\documentstyle[12pt,aasms]{article}
\newcommand{\kms}{km~s$^{-1}$}
\newcommand{\al}{$\alpha$}
\newcommand{\lam}{$\lambda$}

\begin{document}

\title{A New Measurement of the Cosmic Microwave Background Radiation
Temperature at z = 1.97\footnote{Observations here were obtained with the
Multiple Mirror Telescope, a joint facility of the University of Arizona and the Smithsonian Institution.}}

\author{Jian Ge, Jill Bechtold, John H. Black}
\affil{Steward Observatory, University of Arizona, Tucson, AZ 85721}

\begin{abstract}
We present detections of absorption from the ground state  and excited
states of C I in the z = 1.9731 damped Ly\al\ system of the QSO 0013$-$004. 
 The excitation temperature between the J = 0 and J = 1 fine-structure 
levels of C I is 11.6 $\pm$ 1.0 K. 
We estimate other contributions to the excitation of the C I fine-structure
levels, and use the
population ratio of the excited state to the ground state to derive
 an estimate for the  cosmic microwave background radiation (CMBR)
temperature of T = 7.9 $\pm$ 1.0 K at 0.61 mm   and z = 1.9731, which
is consistent with the predicted value of T = 8.105 $\pm$ 0.030
 K  from the standard  cosmology. 

\end{abstract}


\section{Introduction}
The standard Friedman cosmology predicts a simple relationship between the temperature of
 the 
cosmic microwave background radiation (CMBR) and redshift, z:
\begin{equation}
T_{CMBR}(z) = T_{CMBR}(0)(1+z),
\end{equation}
where T$_{CMBR}$(0) is the 
 CMBR temperature today (e.g. Peebles 1993).  The  present-day
 CMBR temperature has 
been measured  precisely with the FIRAS instrument on the Cosmic Background Explorer (COBE), with  T$_{CMBR}$(0) = 2.726$\pm 0.010$ K (at the 95\% confidence level, Mather et al. 1994).

The CMBR temperature at higher redshifts can be measured indirectly by
 using atomic fine-structure 
transitions in absorbers toward high redshift quasars (Bahcall \& Wolf 1968). The first attempt to
measure the CMBR temperature in this way   gave an upper limit for the CMBR temperature, T$_{CMBR} <$ 45 K,  
at z = 2.309 from limits on the  fine-structure excitation 
of C II toward PHL 957 (Bahcall et al. 1973).
 Compared with other abundant species (such as
O I, C II, Si II, N II), C I is a better species to use because 
it has the  smallest
 energy separations in its fine-structure levels. The ground term of C I 
is split into three levels (J = 0, 1, 2) with J = 0 - 1 and J = 1 - 2  
separations 
 of 23.6 K  and 38.9 K (or 0.61 mm and 0.37 mm). 
 Meyer
et al. (1986) used the C I fine structure lines of a damped Ly$\alpha$
system in the spectrum of the QSO 1331+170 to obtain 
an upper limit (2$\sigma$) of T$_{CMBR} < $ 16 K at z = 1.776.
More recently, Songaila et al. (1994b)
 have observed QSO 1331+170 again and obtained
T$_{CMBR}$ = 7.4$\pm 0.8$ K, which agrees with the 
predicted value of 7.58 K. C II is another good species to use for the CMBR 
measurements at high redshift because it has reasonably small energy 
separation between
its fine-structure levels, 91.3 K. Songaila et al. (1994) obtained a 
2 $\sigma$  upper limit of T$_{CMBR} < 13.5$ K at z = 2.909 toward QSO 0636+680
based on upper limits to C II fine-structure. Lu et al. (1995) achieved 
a 3 $\sigma$ upper limit of T$_{CMBR} < 19.6$ K at z = 4.3829 toward QSO 
1202$-$07 by measuring  upper limits for the excited states of C II.

There are several difficulties in carrying out measurements of 
T$_{CMBR}$(z) with quasar absorbers. First, the ground state 
C I absorption lines are 
often  weak and difficult to detect in 
quasar absorbers at high redshift. 
 Second, other non-cosmological sources such as collisions and pumping by UV radiation can also populate  the excited 
fine-structure levels of C I. Thus, the excitation temperature derived is
an upper limit to the CMBR temperature, unless  the local excitation   
can be estimated. Third,
most absorption lines from  abundant species such as O I, C II, Si II, N II 
show strong saturation  in their ground state transitions and
 hence the population ratio of their  excited  state to the ground state cannot
 be accurately determined.

In this paper we present spectra obtained at  the
 Multiple Mirror Telescope (MMT) 
  of C I and C I$^*$ absorption in
the z = 1.9731 damped Ly$\alpha$ system toward the QSO 0013$-$004
and 
estimate the contributions of the various sources of excitation. 
The neutral
 hydrogen column density of the z = 1.9731 damped system is 
N(H I) = $5(\pm 1)\times 10^{20}$ cm$^{-2}$ (Pettini et al. 1994). The metal
abundance is
about 1/4 of the solar value and the heavy element depletion  by dust
 is more than 20\%  of the Milky Way value (Pettini et  al. 1994). These
properties suggested to us that this system was a good candidate for 
 a search for C I absorption.

\section{Observations}

The observations of QSO 0013$-$004 were obtained on October 9 and December
8, 1994 with 
the Blue Channel Spectrograph and  the 
Loral 3072$\times$1024 CCD   on the MMT.
The 832 l/mm grating was used in second order. 
A CuSO$_4$ filter was used to block the first order light.
 In October, we took  three  50-minute and one 60-minute exposures
 with wavelength
coverage from 3860 \AA\ to 4960 \AA. Because of poor seeing conditions, 
a 1.5$''\times 180''$ slit was used to
get a spectral resolution of 1.3 \AA\ (FWHM).  In December,  we
took four 50-minute exposures with wavelength coverage from 4380 \AA\ to 5459 
\AA. 
A 1$''\times 180''$ slit was used to obtain a spectral resolution of 1 \AA\ 
(FWHM). In all our observations, the quasar was moved  a few arcseconds
along the slit between each exposure to smooth out any residual irregularities 
in  the detector response which remained after 
flat-fielding. An exposure of a He-Ne-Ar lamp and a quartz lamp were taken
before and after each exposure of the object to provide an accurate wavelength
reference, a measure of the instrumental resolution, and a flat-field correction.
The spectra were reduced using  standard routines in IRAF, and were summed with 
individual exposures weighted by the signal-to-noise ratio (S/N).
We then summed the spectra with the wavelength coverage from 4550 to 5940 
\AA\ from our two  runs to reach  S/N of about 40.

Figure 1 shows the total spectrum obtained.  All reported wavelengths are 
vacuum and have been corrected to the heliocentric frame.
The continuum was fitted and significant absorption features were identified
and measured  in the way  described by Bechtold (1994). The spectra
shown were  normalized by their fitted continuum. All absorption lines with more than 5 $\sigma$ 
significance  are marked. 
 Table 1  shows the equivalent widths of  the absorption lines
and their identifications.
 The equivalent widths were measured by
specifying start and  stop wavelengths for each absorption feature by hand 
(cf. Bechtold 1994).
The central wavelength of each line  is the centroid, weighted by 
the depth of each pixel in the line profile below the continuum. The error
for the central wavelength shown in this table is from the uncertainty in the
measurement of the equivalent width. 
 There are  at least four velocity components associated with the
z = 1.9731  damped system. The redshifts are z = 1.9673, 1.9700, 1.9714, 
1.9731. Two components (z = 1.9673, 1.9731)  clearly show absorption lines
from the C I ground state levels. Since  some important lines such as C II \lam\
1334  and C I \lam\ 1560 lines are blended  lines, 
 we have also tried to fit the 
absorption lines with Gaussians.
 This method
 is similar to  the method described by Schneider et al. (1993).
  The equivalent widths measured in this way are consistent 
with the ones listed in Table 1
 within the 1 $\sigma$ errors even for  heavily blended
absorption lines. Figure 2 shows the Gaussian fits for the
 C II \lam\ 1334 line and C I \lam\ 1560 line at z = 1.97.

Table 2 lists the rest wavelengths, predicted wavelengths, 
and f-values for the two strongest C I 
multiplets and the strongest C II multiplet in the z = 1.9731
component. The f-values
 are from the compilation of Morton (1991). We also list the observed 
equivalent widths of these lines.
Figure 3 shows our spectrum of QSO 0013$-$004 in the vicinity of the two C I 
multiplets and one C II multiplet 
listed in Table 2. The fit of the continuum and the
1 $\sigma$ deviation of each pixel are also displayed.
C I J = 0  absorption lines are 
 clearly present in  UV multiplet 2 at 1656.93 \AA\ and multiplet 
3  at 1560.31 \AA.  C II J = 1/2 (1334.53 \AA), J = 3/2 (1335.70 \AA)
 absorption lines of multiplet 1 are also present.
  C I J = 1 absorption is present in the
multiplet 2 at 1656.27 \AA\ and 1657.91 \AA\ and multiplet 3 at 1560.68 \AA.
  The C I and C II lines are 
observed at the wavelengths expected from  the
redshift of other low-ionization ions, such as Zn II (Pettini et al. 1994), Fe II, and Si II (Table 1), within  the wavelength uncertainty of about 
0.1 \AA.
 The C I J = 0 absorption line at $\lambda =1656.928$ \AA\
 is blended with one of the 
C I J = 1 lines at $\lambda = 1657.379$ \AA\ and also one
of the C I J = 2 lines at $\lambda = 1657.008 $ \AA.   
 No absorption features for J = 2 at 
$\lambda = 1658.121$ \AA\ and $\lambda\lambda\lambda = 1561.340, 1561.367, 
 1561.438$ \AA\ are detected.
 The third strongest C I multiplet at \lam\ = 1329 \AA\ is blended 
with Si II \lam\ 1304 \AA\ from another absorber at z = 2.0290.


\section{Results}

We can  use the relative population ratios of the J = 1 and J = 0 levels 
in the multiplets
2 and 3  to obtain the excitation temperature of  the C I
 fine-structure levels in its ground state and   to derive limits on
 the CMBR temperature at z = 1.9731.
Since our spectral resolution is insufficient to 
resolve the profiles of the lines,
we  used   observed Si II lines, at \lam\ 1260, \lam\ 1304, \lam\ 1526 and
\lam\ 1808,
 to construct an empirical curve-of-growth (Figure 4). The measurement
 of Si II \lam\
 1206
is from another observation by Bechtold, who found  a
 rest frame equivalent width for 
this line of 0.8691 $\pm$ 0.0188 \AA.
The Si II  curve-of-growth provides a Doppler parameter b = 42 $\pm$ 2 
 \kms\ which we then
used to infer the  column densities of different absorption lines.
The results of calculated column densities are shown in Table 3. We have also
shown  central optical depths for different C I and C II lines. The central
optical depths for C I and C I$^*$ absorption lines indicate that all these
lines are on the linear part of the curve-of-growth. Thus, the derived
column densities for the C I fine structure levels are independent of the 
derived  b value. However, the optical depths for the C 
II  and C II$^*$ lines indicate that they are saturated, and so
the derived column densities for the 
C II and C II$^*$ line depend  on the b-value. 
The derived b-value indicates 
that there are probably several velocity components blended with each other. However, the uncertainties in the column densities from
single b-value curve-of-growth analyses are usually on the order of 
 a factor of 2 (Jenkins
1986). We therefore  use this b-value to derive the column densities for the saturated
 C II and C II$^*$ lines.  In our 
calculation, because of our limited resolution,
 we have  combined the f-values of the two J = 1 lines of C I multiplet 3, 
\lam\ \lam\ 1560.682 \AA, 1560.709 \AA, and
also  the f-values of the two J = 3/2 lines of C II multiplet 1, 
\lam\ \lam\ 1335.663 \AA, 1335.708 \AA, and further derived the relative 
population ratios of their fine structure levels. We have  assumed that
 the absorption at $\lambda = 4926.313$ \AA\
 is only from the J = 0, $\lambda = 1656.928$ \AA\ of C I multiplet 2 since the 
strengths of other blended lines such as J = 1, $\lambda = 1657.379$ \AA\ and 
J = 2, $\lambda = 1657.008$ \AA\ are much weaker than that of the J = 0 
line. The other two J = 1 lines of C I multiplet 2, \lam\ \lam\
 1656.267 \AA, 1657.907 \AA\
are detected  at about the 2-3 $\sigma$ level, and in the weighted mean are 
present at the 4 $\sigma$ level. We therefore
have  used this weighted mean to derive the 
 population ratio of the J = 1 and J = 0 levels of C I multiplet 2, as shown in 
Table 3.

 Next, we can  use  the relative population ratios
to derive the excitation temperature of the C I and C II
 fine-structure levels.
According to the Boltzmann equation, an excitation temperature T$_{ex}$
can be expressed in
terms of the column densities N$_e$ and N$_g$ in the excited  and the 
ground state levels, 
\begin{equation}
N_e/N_g=g_e/g_g~exp(-\Delta~E_{eg}/kT_{ex}),
\end{equation}
  where $\Delta$E$_{eg}$ is the energy 
difference between the excited and ground levels. $\Delta$E$_{eg}$
 is 23.6 K for 
the difference
between J = 1 and J = 0 in C I and 91.2 K for the difference between J = 3/2
and J= 1/2 in C II.  The weights are $g_J = 2J + 1$.
Thus, the  population ratios N(J = 1)/N(J =0) in the C I 
multiplets 2 and 3
indicate  excitation temperatures, T$_{ex}$ = 11.6 $\pm$ 1.6 K and 11.6 $\pm$ 
1.4 K for multiplets 2 and 3, respectively. The weighted 
 mean value is T$_{ex}$ = 11.6  $\pm$ 1.0 K for the C I fine structure.
 The population
ratio N(J = 3/2)/N(J = 1/2) of the C II fine-structure levels
 indicates an excitation temperature, T$_{ex} = 16.1\pm1.4$ K.  
 Because  the C I and C II fine-structure levels can be excited by not only the
CMBR field but also other excitation sources such as collision and 
UV pumping, the derived excitation temperatures are
  upper limits to the CMBR temperature at z = 1.9731. Thus, the upper limits of the CMBR temperature at 0.61 mm and 0.16 mm are 11.6 K and 16.1 K, respectively,
consistent with the predicted value  at this redshift,  T$_{CMBR}$ = 8.105 K.

  We can 
estimate the  contribution    from  collisional
and UV pumping to the excitation of C I by modeling the absorption region.  The equilibrium between the excitation and 
de-excitation of the C I J = 0$\rightarrow$1 fine structure can be expressed as
\begin{equation}
N_0[\sum_{j}<\sigma_{01} v>_{j} n_j + B_{01}~I_\nu + \Gamma_{01}] = N_1[A_{10} +B_{10}~I_\nu +\sum_{j}<\sigma_{10} v>_{j} n_j + \Gamma_{10}],
\end{equation}
 where j = H, e, p, He and H$_2$, $\Gamma_{10}$ is the 
UV pumping rate from J = 0 to J = 1, the $\Gamma_{01}$ is the UV pumping rate 
from J = 1 to J = 0.
 The spontaneous transition probability for the C I J = 1$\rightarrow$0 
transition $A_{10}=7.93\times 10^{-8}$ s$^{-1}$ (Bahcall \& Wolf, 1968). The 
collisional excitation rates due to different collision partners are given
by Launay et al. (1977); Keenan et al. (1986); 
Johnson et al. (1987); Roueff et al. (1990); Flower (1990); 
Staemmler et al. (1991)
 and Schr\"oder et al. (1991). The collisional de-excitation rate  
\begin{equation}
<\sigma_{10} v>_{j}=1/3<\sigma_{01} v>_{j}exp(23.6~K/T_k),
\end{equation}
 for j = H, e, p, He and H$_2$. $B_{10}=1/3~B_{01}$.  The J = 0$\rightarrow$1 excitation rate due to the absorption of the
CMBR, $B_{01}$, can be expressed as
\begin{equation}
I_\nu B_{01} = 2.38\times 10^{-7}/[exp(23.6~K/T_{CMBR})-1]~s^{-1}. 
\end{equation}
The UV pumping rate 
 depends on the strength of UV radiation field. $\Gamma_{01} = 7.55\times 10^{-10}$ s$^{-1}$ and $\Gamma_{10} = 2.52\times 10^{-10}$ s$^{-1}$ if the UV field intensity
in the z = 1.9731 is  the same  as that in the Milky Way which is  about
 $4.7\times 10^{-19}$  ergs s$^{-1}$ cm$^{-2}$ Hz$^{-1}$ at 912 \AA\
(Jenkins \& Shaya 1979; Mathis et al. 1983).

In order to solve  Eq. (3), we have to know n$_H$, n$_e$, n$_{He}$,
n$_{H_2}$ and the UV pumping rates. To estimate plausible values for the z = 1.9731
 absorber  we constructed a photoionization model with the CLOUDY program (Ferland 
1993). For the input
to CLOUDY, we adopted a metallicity of 25\% of the solar value, i.e.,
[Zn/H] = $-$0.61, for all
the elements in the z = 1.9731 damped system (Pettini et al. 1994).
 We have also considered   depletion by dust grains. The dust-to-gas
ratio is about 20\% of the Milky Way, estimated  from
 the relative depletion of Cr and Ni to Zn, i.e. [Cr/Zn] $\le$ $-$1.15 (Pettini et al. 1994) and [Ni/Zn] 
$\le$ $-$0.98 from our data. For the  
shape of the spectral energy distribution (SED),
 we adopted a parameterization of   the Milky Way  SED given by Black (1987). 
 Because we are interested in the low ionization species (C I, C II and H$_2$), the results are sensitive to the UV flux adopted at  wavelengths
 from $\sim 500-1100$ \AA\ which is probably dominated by local sources within 
the galaxy. The adopted flux at the  Lyman limit 
is about one order of magnitude higher than the metagalactic UV flux
 at z $\approx$ 2, estimated to be 
J(912 \AA) $\approx 3.8 \times 10^{20}$ ergs  s$^{-1}$ cm$^{-2}$ Hz$^{-1}$ 
(e.g. Bechtold 1994), so we have neglected the ionization contribution  
from the metagalactic radiation field.  The results
are shown in Figure 5. 
The ionization parameter, U =  $\phi(H)/n_H c = 2.7\times 10^{-4}$,
 gives the best  fit to the 
observational results, where $\phi(H)$ is the surface flux  of 
hydrogen-ionizing photons (cm$^{-2}$ s$^{-1}$). The photoionization model with 
this U-value indicates that
the column density of molecular hydrogen, N(H$_2) = 5\times 10^{19}$ 
cm$^{-2}$, or n$_{H_2} = 0.1 n_H$; the electron temperature, T$_e \sim 1\times 10^3$ - 80 K and
 n$_e$/n$_H$ $\sim 1.0\times 10^{-2}$ - $5\times 10^{-4}$  from the outer region
to the inner region  of the absorber. T$_e$  of $\sim$ 100 K and 
n$_e$/n$_H$ of $\sim$ $5.0\times 10^{-4}$ dominate most regions of the cloud. 
In the following 
discussions we adopted  two sets of extreme limit
values: n$_e$/n$_H = 1.0\times 10^{-2}$ and T$_e$ = 1000 K; n$_e$/n$_H$ =  $5.0\times 10^{-4}$
and T$_e$ = 100 K. 

In order to estimate  n$_H$ we use derived from the relative population
 ratio  of the C II fine-structure levels.
In the H I dominant region with n$_H \la 3\times 10^3$ cm$^{-3}$ (Flower 1990;
Bahcall \& Wolf 1968), 
the 
ratio of excited C II$^*$ relative to the ground state C II populations can be 
expressed as
\begin{equation}
\frac{N(C II^*)}{N(C II)} \approx \frac{n_H <\sigma_{01} v>_{H} + 
n_e<\sigma_{01} v>_{e} + n_{H_2}<\sigma_{01} v>_{H_2}}{A_{10}},
\end{equation}
where $A_{10} = 2.29\times 10^{-6}$ s$^{-1}$  is the spontaneous transition
probability,
 and where we have neglected the excitation term due to 
proton collisions because this term is much less than the others at T$_e <
2\times 10^5$ K (Bahcall \& Wolf 1968). 
As a result, the neutral hydrogen density, n$_H = 21.0\pm 9.6$ cm$^{-3}$ when 
T$_e = 100$ K, and n$_H = 4.5\pm2.0$ cm$^{-3}$ when T$_e = 1000$ K. Thus, the 
H-ionization photon flux $\phi(H)$, is $17.0\times 10^7$ cm$^{-2}$s$^{-1}$ when
T$_e =100$ K, and is $ 3.6\times 10^7$ cm$^{-2}$s$^{-1}$ 
when T$_e = 1000$
K. For comparison, the Milky Way H-ionization flux is about $1\times 10^7$ cm$^{-2}$s$^{-1}$ calculated from the SED  given by Black (1987). So, the  UV pumping rates in our calculations are 17.0 and 3.6 times of the 
Milky Way rate for the 100 K and 1000 K cases, respectively. We obtain
 n$_e$ = 1.05$\times 10^{-2}$ cm$^{-3}$, n$_{He}$ = 1.68 cm$^{-3}$ and 
n$_{H_2}$ = 2.1 cm$^{-3}$ for the T$_e$ = 100 K case and n$_e$ = 4.5$\times 
10^{-2}$ cm$^{-3}$, n$_{He}$ = 0.36 cm$^{-3}$ and n$_{H_2}$ = 0.45 cm$^{-3}$
for the T$_e$ = 1000 K case.

Substituting  into  Eq. (4), we finally  estimate the
 contribution to the excitation of the C I fine structure levels from  
collisions and UV pumping.
 After these  contributions are removed,  the CMBR temperature at
 z = 1.9731 T$_{CMBR} = 7.9 \pm 1.0$ K when
T$_e = 100$ K is adopted, and T$_{CMBR} = 10.6 \pm 1.0$ K when T$_e = 1000$ K
is used. Since  the electron temperature 
in most regions of the z = 1.9731 absorber is around 100 K, 
 our best guess for the CMBR temperature at z = 1.9731 is  7.9 $\pm$ 1.0 K.

The above results are based on the assumption of a single homogeneous zone 
model, which is probably 
different from the real case. Previous high resolution 
observations of the QSO 1331 + 170 have shown that the C I absorption lines split
into two components with different excitation temperature (Songaila et al. 
1994). There may be  two or  more different velocity components
associated with C I absorption in the QSO 0013$-$004 system. 
Without  knowledge of the individual cloud structure, there may be
some uncertainties in the correction of local excitation from only considering
the  C II fine structure excitation.
Ultimately, a higher resolution spectrum
 is needed to get an improved
  measurement of the CMBR temperature at  z = 1.9731.

\section{Discussion}

 We have estimated the   local contributions to the  
excitation of C I , which can contribute
$\sim 1-3$ K  to the excitation temperature of the  C I ground state
fine-structure levels at z $\sim$ 2
 in  reasonable physical and chemical conditions for C I to exist. 
 After estimating these local contributions, our best guess
 for the CMBR temperature is 7.9 $\pm 1.0$ K, which
 is consistent with the predicted value of 8.105 K, at z  = 1.9731.

Our study  shows  that the local contributions to the 
excitation of the C I fine-structure levels  are dominated 
by  collisions with neutral hydrogen  and UV pumping. 
However, if the number density of molecular
hydrogen is comparable to that of neutral hydrogen, H$_2$ can also be 
an important collisional partner for C I excitation. 
At high electron temperature (e.g. 1000 K or higher) electrons can
also be  important collisional partners.

Our study also shows that the UV radiation field in the z = 1.9731 absorber
is about 10 times stronger than the average value in the Milky Way.
This could be the result of a higher star formation rate in this system.

Figure 6 illustrates the 
 measurements of  the CMBR temperature at different redshifts. 
All high redshift measures are essentially upper limits, since 
local contributions to the
C I and C II excitation may be significant. So far, 
all measurements are consistent with
the Big Bang predictions.

\acknowledgements
We thank Dr. A. Songaila  for pointing out an important
point which improved the paper.
We  thank G. Ferland for providing  his CLOUDY program.
 We also thank the staff of MMTO for their  help. 
This research was supported by NSF AST-9058510 and NASA grant NAGW-2201.

\newpage
\centerline{\bf Figure Captions}

Figure 1.---Spectrum of QSO 0013$-$004 with significant absorption lines marked. The solid line is the fit to the
 continuum. The dotted line is the 1 $\sigma$ error. The  features marked
with an asterisk 
are bad columns or traps in the CCD.

Figure 2.---(a). Profile fit to the C II lines at z = 1.9672, 1.9711 and 1.9733, and
C II$^*$ line at z = 1.9732 . The solid line is the continuum. The dot-dashed line is 
the 1 $\sigma$ errors. The dashed line is the 
Gaussian fit to the absorption lines with four velocity components (see Table 1 and text). 
(b). Profile fit to the C I \lam\ 1560.31 \AA\ and C I$^*$ \lam\ \lam\
 1560.68, 1560.70 \AA\ lines at z = 1.9731. The solid line is the continuum.
The dashed line is the Gaussian fit to the absorption lines.

Figure 3. ---Spectrum of  QSO 0013$-$004 showing the C I multiplets
2 and 3  and C II multiplet 1 listed in Table 1.
 The dotted line shows the 1 $\sigma$ error.
The expected positions of the ground state (J = 0 or 1/2) and excited states
 (J =1, 2 or 3/2)
are marked. 

Figure 4.---Curve of growth for singly ionized and neutral species
 in the z = 1.9731 system. The 
solid line which best fits the data from Si II absorption lines is the
theoretical curve of growth for b = 42 $\pm$ 2 \kms.

Figure 5.---Results   of ionization model for a H I dominant region with N(H I)
= $5\times 10^{20}$ cm$^{-2}$. The ordinate is the column density of 
various ions, the abscissa is the log of ionization parameter U. The solid
lines are predicted values from  CLOUDY. The dotted lines correspond
 to measured column
densities of C II and C I.
The error bars are  
 1 $\sigma$ errors on the column densities of C I and C II which
include errors from photon statistics and uncertainty in the b-value 
(42 $\pm$ 2 \kms).

Figure 6.---Measurements of the CMBR temperature as a function of redshifts.
The
solid line  is the predicted relation. The filled circle
is  from the COBE measurement (Mather et al. 1994). The open squares are upper limits obtained
by Songaila et al. (1994a,b). The filled square is obtained here.
The filled hexagon is obtained by
Lu et al. (1995).

\newpage 

\textwidth 230mm
\textheight 280mm
\topmargin -10 mm
\oddsidemargin 5mm
\renewcommand{\baselinestretch}{0.9}
\rm
\scriptsize
\begin{flushleft}
Table 1. The identifications of absorption lines  of QSO 0013-004
\end{flushleft}
\begin{tabular}{cccccccclc}  \hline
No. & $\lambda_{obs}$(\AA) & $\sigma(\lambda)$ & W$_{obs}$ (\AA) & $\sigma(W)$&
W$_{obs}$ (\AA)$^a$&$\sigma(W)^a$ & Significance Level& ID & z$_{abs}$\\ 
\hline
 &&&&&&&\\
1&3869.22&0.03&2.145&0.081&&&26.62&O I 1302&1.9714\\
2&3871.58&0.03&2.719&0.093&&&29.28&O I 1302&1.9732\\
&&&&&&&&Si II 1304& 1.9673\\
3&3874.02&0.03&0.537&0.057&&&9.41&Si II 1304&1.9700\\
&&&&&&&&O I 1302&1.9756\\
4&3875.63&0.04&1.336&0.074&1.22&0.23&18.16&Si II 1304&1.9713\\
&&&&&&&&O I 1302&1.9776\\
5&3878.03&0.05&1.974&0.086&2.12&0.22&22.31&Si II 1304&1.9731\\
6&3936.93&0.09&0.564&0.055&&&10.32&C IV 1548&1.5429\\
7&3943.25&0.05&0.286&0.037&&&7.79&C IV 1550&1.5428\\
8&3944.27&0.09&0.257&0.041&&&6.23&O I 1304&2.0290\\
9&3951.09&0.08&0.252&0.040&&&6.28&Si II 1304&2.0291\\
&&&&&&&&C I 1329&1.9731\\
10&3959.78&0.02&2.853&0.051&2.91&0.19&55.44&C II 1334&1.9672\\
11&3964.48&0.02&4.581&0.052&4.80&0.70&88.57&C II 1334&1.9711\\
12&3967.95&0.02&2.664&0.042&2.31&0.58&63.68&C II 1334&1.9733\\
13&3971.34&0.05&0.797&0.045&0.80&0.09&17.81&C II$^*$ 1335&1.9732\\
14&4042.49&0.06&0.855&0.046&&&18.54&C II 1334&2.0292\\
15&4135.63&0.07&0.730&0.046&&&15.80&Si IV 1393&1.9674\\
16&4139.41&0.03&1.137&0.039&&&29.10&Si IV 1393&1.9700\\
17&4141.78&0.02&1.577&0.041&&&38.54&Si IV 1393&1.9717\\
18&4162.39&0.10&0.406&0.043&&&9.34&Si IV 1402&1.9673\\
19&4166.10&0.04&0.654&0.039&&&16.99&Si IV 1402&1.9699\\
20&4168.29&0.04&1.031&0.043&&&23.97&Si IV 1402&1.9715\\
21&4201.21&0.02&1.961&0.040&&&48.43&C IV 1548&1.7136\\
22&4208.07&0.03&1.610&0.043&&&37.07&C IV 1402&1.7135\\
23&4221.58&0.03&1.045&0.036&&&29.02&Si IV 1393&2.0289\\
24&4248.90&0.08&0.578&0.044&&&13.20&Si IV 1402&2.0290\\
25&4530.40&0.02&1.980&0.043&&&46.48&Si II 1526&1.9674\\
26&4534.19&0.06&0.478&0.038&&&12.53&Si II 1526&1.9699\\
27&4536.38&0.02&1.754&0.038&1.89&0.14&46.13&Si II 1526&1.9714\\
28&4538.98&0.02&2.486&0.042&2.55&0.14&66.14&Si II 1526&1.9731\\
29&4572.32&0.11&0.146&0.026&&&5.59&&\\
30&4587.70&0.16&0.158&0.031&&&5.11&&\\
31&4594.00&0.03&1.880&0.037&&&51.47&C IV 1548&1.9673\\
32&4597.70&0.02&1.746&0.030&&&58.65&C IV 1548&1.9697\\
33&4601.13&0.02&3.724&0.039&&&95.74&C IV 1548&1.9719\\
34&4605.54&0.01&1.804&0.027&&&66.51&C IV 1550&1.9698\\
35&4608.24&0.01&2.396&0.031&&&77.52&C IV 1550&1.9716\\
&&&&&&&&C IV 1548&1.9756\\
36&4610.95&0.04&0.515&0.026&&&19.60&C IV 1550&1.9733\\
&&&&&&&&C IV 1548&1.9777\\
37&4614.64&0.05&0.592&0.032&&&18.45&C IV 1550&1.9757\\
38&4617.95&0.08&0.250&0.029&&&8.72&C IV 1550&1.9778\\
39&4630.00&0.015&0.162&0.030&&&5.31&C I 1560&1.9674\\
40&4638.81&0.08&0.198&0.026&0.17&0.02&7.57&C I 1560&1.9730\\
41&4689.12&0.03&1.760&0.036&&&49.04&C IV 1548&2.0288\\
42&4696.79&0.03&1.292&0.032&&&40.91&C IV 1550&2.0287\\
43&4772.77&0.06&0.372&0.018&&&20.73&Fe II 1608&1.9673\\
44&4777.48&0.06&0.074&0.011&&&7.04&Fe II 1608&1.9702\\
45&4779.34&0.02&0.548&0.013&&&41.00&Fe II 1608&1.9714\\
46&4782.21&0.02&0.992&0.016&&&60.25&Fe II 1608&1.9731\\
47&4916.83&0.21&0.321&0.047&&&6.77&C I 1656&1.9674\\
48&4926.08&0.10&0.432&0.043&0.44&0.05&10.13&C I 1656&1.9730\\
49&4957.65&0.05&1.531&0.066&&&23.29&Al III 1670&1.9673\\
50&4962.44&0.08&0.923&0.062&&&14.83&Al III 1670&1.9701\\
51&4964.81&0.02&1.753&0.048&&&36.25&Al III 1670&1.9715\\
52&4967.65&0.05&2.232&0.067&&&33.11&Al III 1670&1.9732\\
53&5061.16&0.15&0.425&0.061&&&6.94&Al III 1670&2.0292\\
54&5208.79&0.16&0.735&0.091&&&8.12&&\\
55&5211.96&0.17&0.523&0.085&&&6.16&&\\
&&&&&&&\\
\hline

\end{tabular}

$^a$The eqivalent widths are measured through Gaussian profile fitting.


\newpage

\textwidth 250mm
\textheight 260mm
\topmargin -10 mm
\oddsidemargin -5mm
\renewcommand{\baselinestretch}{1.1}
\rm
\begin{flushleft}

Table 2. The expected wavelengths for C I and C II in the z = 1.9731 absorber 

\end{flushleft}

\begin{tabular}{cclccllc}  \hline\hline
Multiplet& J&$\lambda_{rest}$(\AA) & $\lambda_{rest}(1+z)$(\AA)$^a$&f&W$_{obs}$(\AA)&$\sigma(W)$&SL\\ 
\hline
 &&&&&&&\\
C I multiplet 2&0&1656.928&4926.213&0.141&0.432&0.043&10.1\\
 &1&1656.267&4924.247&0.059&0.088&0.027&3.3\\
 &1&1657.379$^b$&4927.554&0.035&&&\\
 &1&1657.907&4929.123&0.047&0.054&0.028&1.9\\
 &2&1657.008$^b$&4926.451&0.105&&&\\
 &2&1658.121$^c$&4929.594&0.035&$<$0.111&&\\
 &&&&&&&\\
C I multiplet 3&0&1560.309&4638.955&0.080&0.198(0.17)$^f$&0.026(0.02)$^f$&7.6\\
 &1&1560.682&4639.908&0.060&0.082(0.09)$^f$&0.019(0.02)$^f$&4.4\\
 &1&1560.709$^d$&4640.144&0.020&&&\\
 &2&1561.340$^c$&4642.020&0.012&$< 0.087$&&\\
 &2&1561.367$^c$&4642.100&0.001&$<0.087$&&\\
 &2&1561.438$^c$&4642.311&0.068&$<0.087$&&\\
&&&&&&&\\
C II multiplet 1&1/2&1334.532&3967.698&0.128&2.664&0.039&68.23\\
&3/2&1335.663$^e$&3971.059&0.013&&&\\
&3/2&1335.708&3971.193&0.115&0.797&0.042&18.92\\
&&&&&&&\\
\hline

\end{tabular}

$^a$These wavelengths are vacuum, heliocentric values. 

$^b$The line is blended with J = 0 line of $\lambda = 1656.928$ \AA.

$^c$The upper limits are 3 $\sigma$.

$^d$The line is blended with J = 1 line of $\lambda = 1560.709$ \AA.

$^e$The line is blended with J = 3/2 line of $\lambda = 1335.708$ \AA.

$^f$The measurement in the bracket is from the Gaussian profile fitting.

\newpage

\textwidth 80mm
\textheight 260mm
\topmargin -10 mm
\oddsidemargin -5mm
\rm
\begin{flushleft}

~~~Table 3. Excitation  Temperature of C I and C II at   z = 1.9731.

\end{flushleft}

\begin{tabular}{lll}  \hline\hline
 &\\
C I multiplet 2 &&$\tau_0$\\
~~N($\times 10^{13}$ cm$^{-2}$):&&\\
~~~~~~J = 0, $\lambda = 1656.928$ \AA\ & 4.9 $\pm$ 0.5&0.43\\
~~~~~~J = 1, $\lambda = 1656.267$ \AA\ & 2.1 $\pm$ 0.6&0.078\\
~~~~~~J = 1, $\lambda = 1657.907$ \AA\ & 1.6 $\pm$ 0.8&0.047\\
~~~~~~$<$J = 1$>_{Weighted}$&1.9 $\pm$ 0.5&\\
~~N(J = 1)/N(J = 0) & 0.39 $\pm 0.11$&\\
~~T$_{ex}$(K)& 11.6 $\pm$ 1.6&\\
&\\
C I multiplet 3 &&\\
~~N($\times 10^{13}$ cm$^{-2}$):&&\\
~~~~~~J = 0, $\lambda = 1560.309$ \AA\ & 4.1 $\pm$ 0.5&0.19\\
~~~~~~J = 1, $\lambda = 1560.695$ \AA\ & 1.6 $\pm$ 0.4&0.076\\
~~N(J = 1)/N(J = 0) & 0.39 $\pm$ 0.10&\\
~~T$_{ex}$(K)& 11.6 $\pm$ 1.4&\\
&\\
Weighted Mean of C I Multiplets &&\\
~~N(J = 0)($\times 10^{13}$ cm$^{-2}$)& 4.5 $\pm$ 0.4&\\
~~N(J = 1)($\times 10^{13}$ cm$^{-2}$)& 1.8 $\pm$ 0.3&\\
~~N(J = 1)/N(J = 0)& 0.39 $\pm$ 0.07&\\
~~T$_{ex}$(K)& 11.6 $\pm$ 1.0&\\
 &\\
C II multiplet 1 &&\\
~~N(J = 1/2)($\times 10^{16}$ cm$^{-2})$& 2.7 $\pm$ 1.2&164\\
~~N(J = 3/2)($\times 10^{14}$ cm$^{-2})$& 1.9 $\pm$ 0.2&1.1\\
~~N(J = 3/2)/N(J = 1/2)& 7.0($\pm$ 3.2)$\times 10^{-3}$&\\
~~T$_{ex}$(K)& 16.1 $\pm$ 1.4&\\
&\\ 
\hline

\end{tabular}
\normalsize

\newpage
\begin{figure}
\plotone{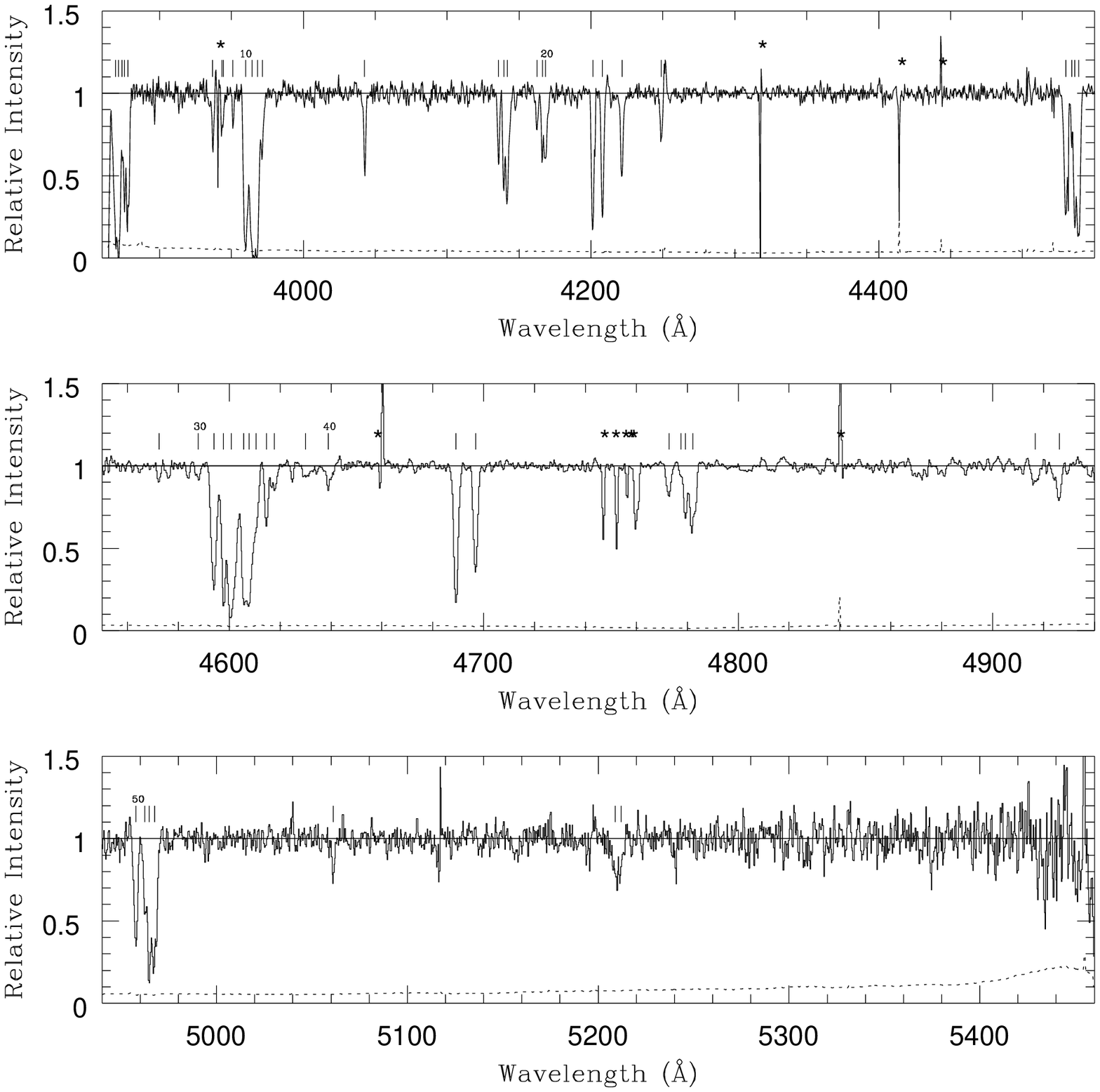}
\end{figure}

\newpage
\begin{figure}

\plotfiddle{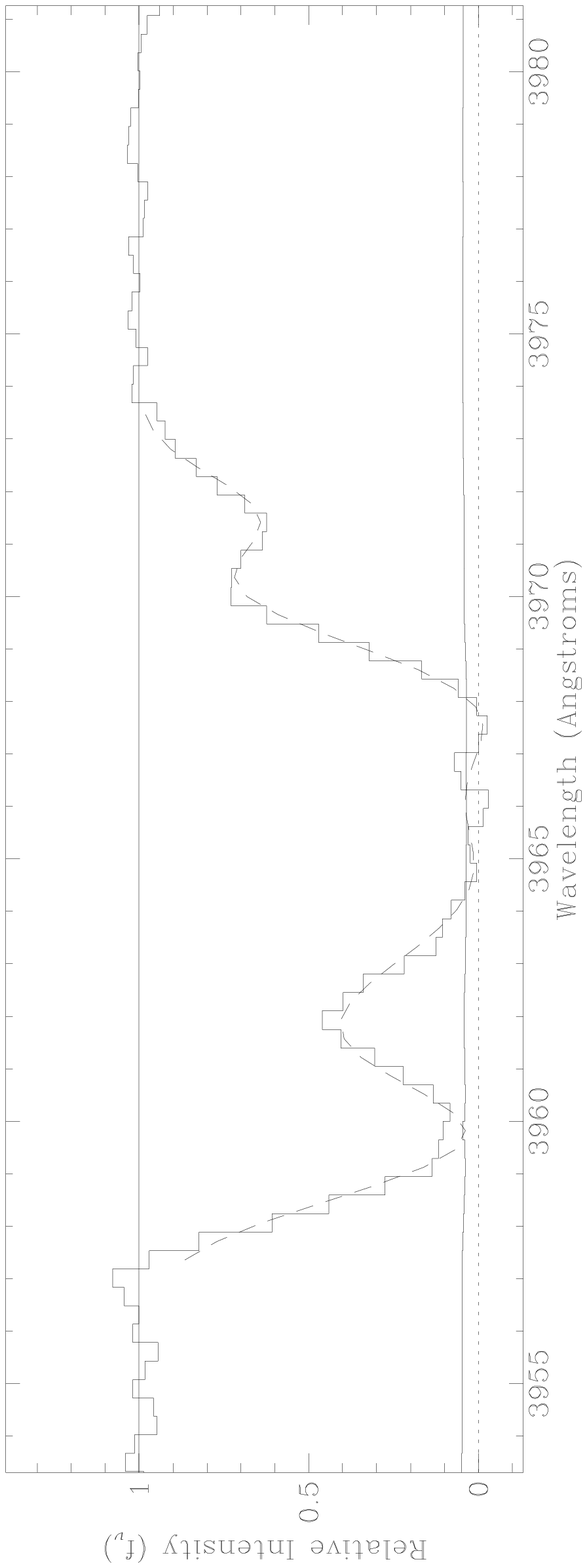}{40mm}{-90}{60}{60}{-240}{350}
\plotfiddle{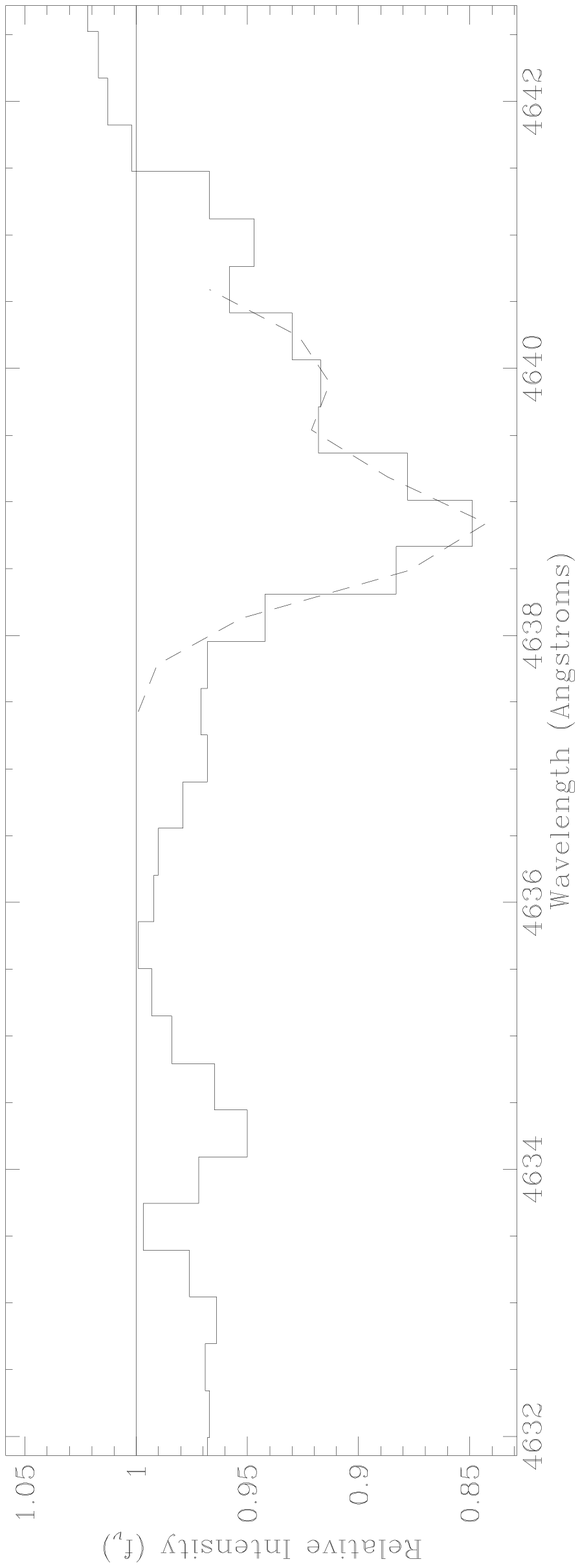}{40mm}{-90}{60}{60}{-240}{300}
\end{figure}

\newpage
\begin{figure}
\plotone{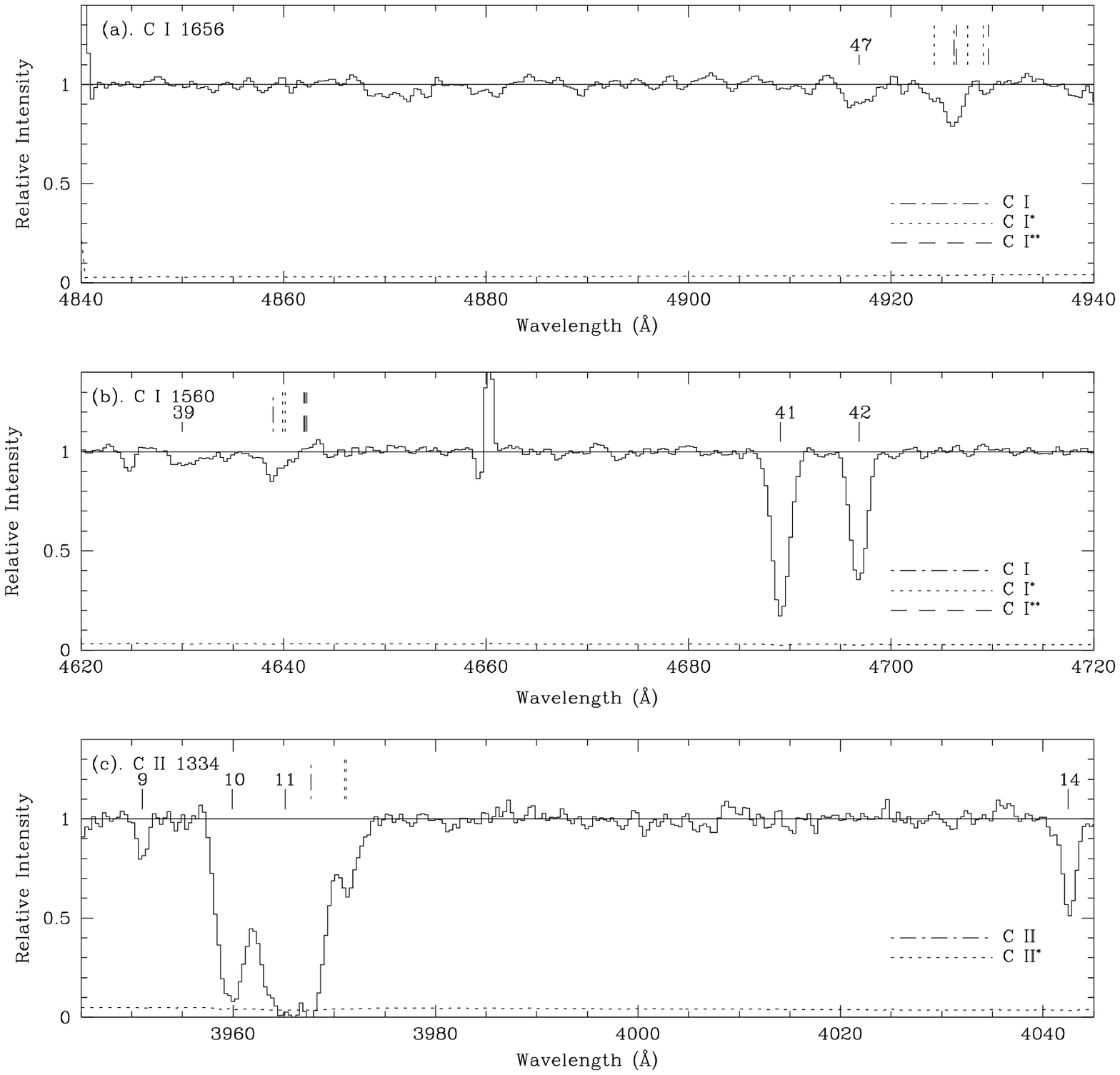}
\end{figure}

\newpage
\begin{figure}
\plotone{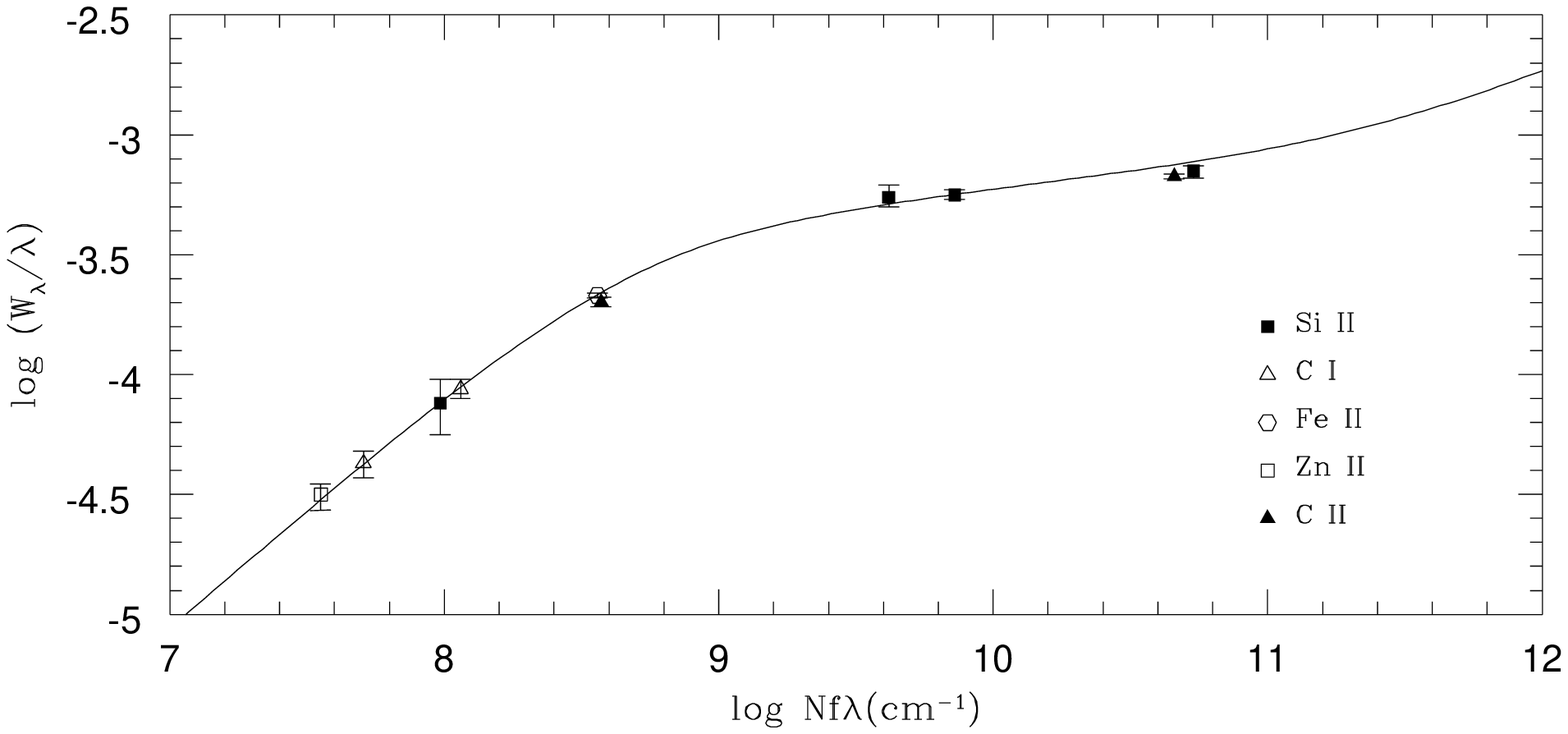}
\end{figure}

\newpage
\begin{figure}
\plotone{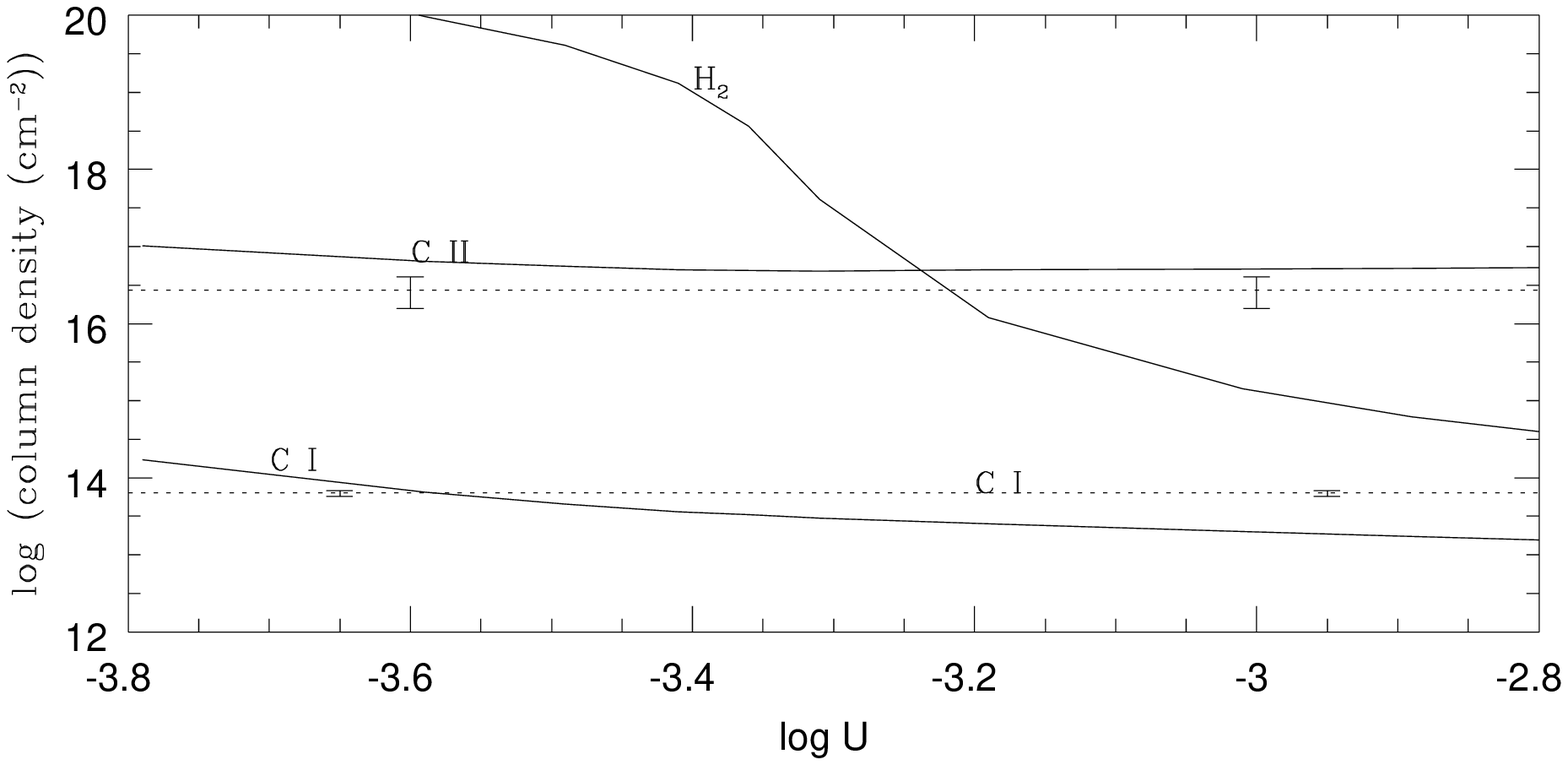}
\end{figure}

\newpage
\begin{figure}
\plotone{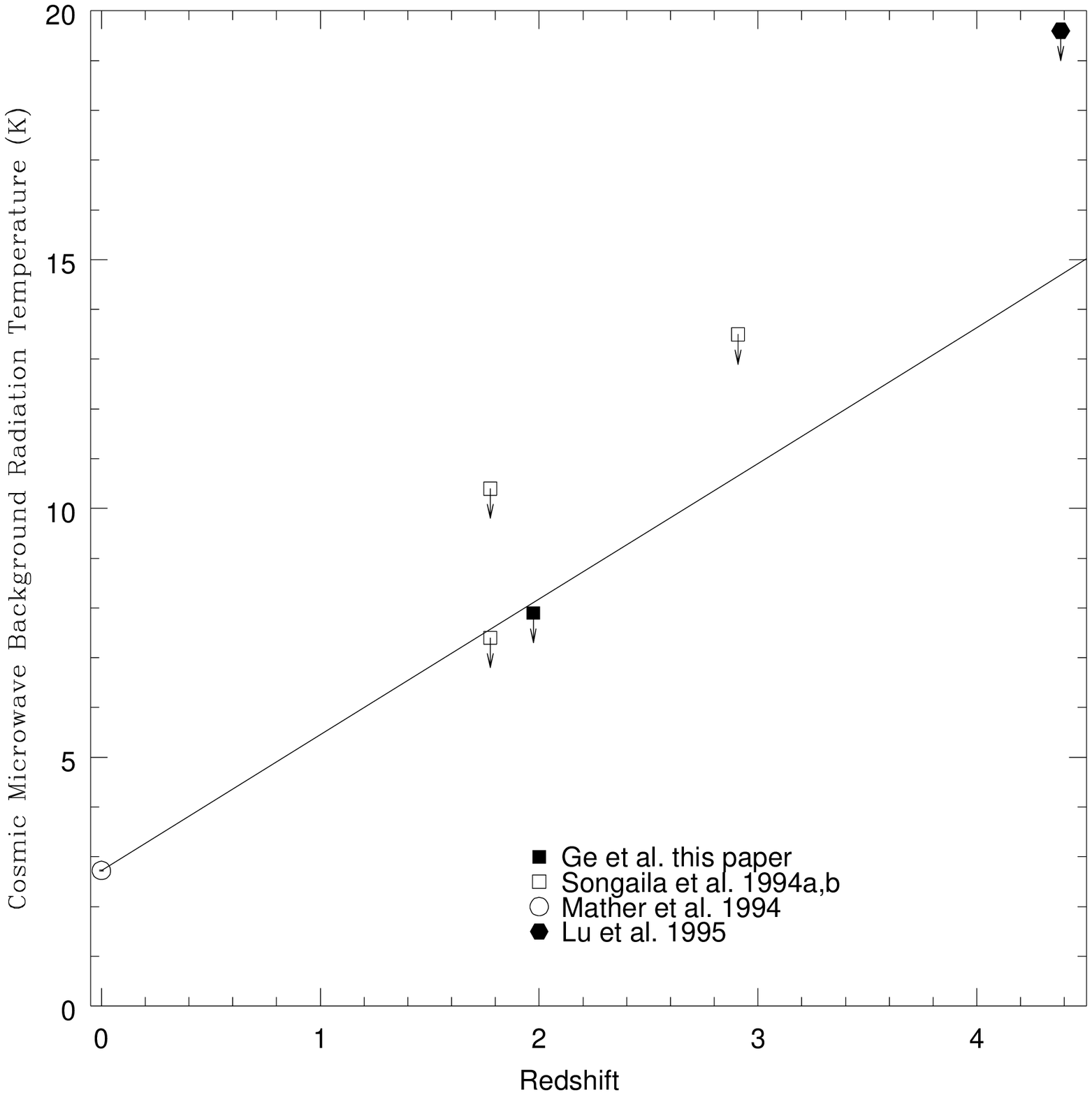}
\end{figure}

\end{document}